\documentclass[journal,onecolumn]{IEEEtran}
\usepackage{amsmath}
\usepackage{amsfonts}
\usepackage{multirow}
\usepackage{graphicx}
\ifCLASSINFOpdf
\else
\fi
\begin{document}
%
\title{Statistical Analytics and Regional Representation Learning for COVID-19 Pandemic Understanding}
%
%
%

\author{Shayan Fazeli,~\IEEEmembership{Ph.D.,~UCLA,}
        Babak Moatamed,~\IEEEmembership{Ph.D.,~UCLA,}
        and~Majid~Sarrafzadeh,~\IEEEmembership{Ph.D.,~UCLA}
}

%
%

\markboth{June 9th, 2020}%
{Shell \MakeLowercase{\textit{et al.}}: Bare Demo of IEEEtran.cls for IEEE Journals}
%



\maketitle

\begin{abstract}
\textbf{Background: } The rapid spread of the novel coronavirus (COVID-19) has severely impacted almost all countries around the world. It not only has caused a tremendous burden on healthcare providers to bear, but it has also severely impacted the global economy and social life. The presence of reliable data and the results of in-depth statistical analyses provide researchers and policymakers with invaluable information to understand this pandemic and its growth pattern more clearly.

\textbf{Objective:} This study aims to gather an extensive collection of fine-grained regional features along with COVID-19 pandemic patterns across the United States and to design statistical methodologies and machine learning pipelines to provide means for a more thorough understanding of the underlying patterns and enable the use of Artificial Intelligence.

\textbf{Methods:} This paper processes and combines an extensive collection of publicly available datasets to provide a unified information source for representing areas and sub-regions with regards to their pandemic-related behavior. The features are grouped into various categories to account for their impact based on the higher-level concepts associated with them. This work uses several correlation analysis techniques to observe value and order relationships between features, feature groups, and COVID-19 occurrences. Dimensionality reduction techniques and projection methodologies are used to elaborate on individual and group contribution of these informative features to the representation variance. In addition, a specific RNN-based inference pipeline called DoubleWindowLSTM-CP is designed in this work for predictive event modeling with minimal use of historical data on outbreak events, thus utilizing sequential and temporal patterns as well as enabling concise record representation.

\textbf{Results: } The primarily quantitative results of our statistical analytics indicated critical patterns reflecting on many of the expected collective behavior and their associated outcomes. As an example, the $33\%$ Pearson correlation with a p-value smaller than $0.0001$ indicates a well-defined relationship between the proportion of public transit in the methods of commute to work and the daily number of deaths due to COVID-19. Regarding deep learning, our DoubleWindowLSTM-CP instance with the time window of $t=10$ days exhibits clear training convergence and efficient prediction results.

\textbf{Conclusions:} Representing a region and its population and community can play an essential role in pandemic modeling, and this is due to the fact that such representation reflects on the regional reaction and susceptibility to the outbreak. The analysis presented here demonstrates that high-resolution region-based features can be leveraged to obtain accurate outbreak event predictions while using but a minimal amount of historical data on the pandemic patterns.

\end{abstract}


%
\IEEEpeerreviewmaketitle

\section{Introduction}
%
%
%
%
\IEEEPARstart{I}{n} the early days of the year 2020, the world faced another widespread pandemic, this time of the COVID-19 strand, otherwise known as the novel coronavirus. The family of Coronaviruses to which this RNA virus belongs can cause respiratory tract infections of various severities. These infections range from cases of the common cold to the more lethal degrees. Many of the confirmed cases and deaths reported due to COVID-19 showed evidence of severe forms of the aforementioned infections \cite{king2012virus,fan2019bat,liu2019viral}.
The origin of this new virus is still not clearly understood; however, it is believed to be mainly connected to the interactions between humans and particular animal species \cite{liu2019viral}.

The rapid spread of this virus has led to many lives being lost and extremely overwhelmed the health-care providers. It also led to worldwide difficulties and had considerable negative economic impacts. It is also expected to have adverse effects on mental health as well due to prolonged shutdowns and quarantines, and there are guidelines published to help minimize this negative impact \cite{anxietycovid}.

In this work, we have gathered, processed, and combined several well-known publicly available datasets on the COVID-19 outbreak in the United States. The idea is to provide a reliable source of information derived from a wide range of sources on important features describing a region and its population from various perspectives. These features primarily have to do with demographics, socio-economic, and public health aspects of the US regions. They are chosen in this manner because it is plausible to assume that they can be potential indicators of commonalities between the affected areas. Even though finding causality is not the objective of this work, our analyses attempt to shed light on these possible commonalities that allow public health researchers to obtain a better perspective on the nature of this pandemic and the potential factors contributing to a slower outbreak. This is vitally important as the critical role of proper policies enforced at the proper time is evident now more than ever.

There has been widespread attention in the design and utilization of Artificial Intelligence-based tools to obtain a better understanding this pandemic. Accordingly, we present a neural architecture with recurrent neural networks in its core to allow the machine to learn to predict pandemic events in the near future, given a short window of historical information on static and dynamic regional features. The main assumption that this work attempts to empirically validate is that the concise pandemic-related region-based representations can be learned and leveraged to obtain accurate outbreak event prediction with only minimal use of the historical information related to the outbreak. Aside from the theoretical importance, an essential application of this framework is when the reported historical pandemic information, e.g., number of cases, is not reliable. An example of this is when a region discovers a problem in its reporting scheme that makes the historical information on the pandemic inaccurate due to overestimation or underestimation. Such unreliability will severely affect the models which use this historical information as the core of their analysis.

In summary, the contributions of this work are as follows:
\begin{itemize}
    \item 
Gathering and providing a thorough collection of datasets for the fine-grained representation of US counties as sub-regions. This collection includes data from various US bureaus, health organizations, the Center for Disease Control and Prevention, and COVID-19 epidemic information.
\item Evaluation of the informativeness of individual features in distinguishing between regions
\item Correlation analyses and investigating monotonic and non-monotonic relationships between several key features and the pandemic outcomes
\item Proposing a neural architecture for accurate short-term predictive modeling of the COVID-19 pandemic with minimal use of historical data by leveraging the automatically learned region representations
\end{itemize}

Given the importance of open-research in dealing with the COVID-19 pandemic, we have also designed OLIVIA \cite{oliviaweb}. OLIVIA is our online interactive platform with various utilities for COVID-19 event monitoring and analytics, which allows both expert researchers and users with little or no scientific background to study outbreak events and regional characteristics. The codes for this work and the collection of datasets are also available as well.

\section{Related Works}
Since the beginning of the COVID-19 pandemic, there have been efforts in utilizing computerized advancements in controlling and understanding this disease. An example is the applications developed to monitor the patients' locations and routes of movement. A notable work in this area is MIT's SafePaths application \cite{safepaths} that contains interview and profiling capability for places and paths. It is worthwhile to mention that these platforms have also caused worries regarding maintaining patients' privacy \cite{raskar2020apps}.

To provide researchers and government agencies with frequently updated monitoring information regarding the coronavirus, 1point3acres team has provided an API that allows access to the daily updated numbers of coronavirus cases \cite{covidapi,yang2020covidnet}. Several datasets such as \cite{kaggle10} are also released to the public.

A large corpus of scientific articles on coronaviruses is released as well as a result of a collaboration between AllenAI Institute, Microsoft Research, Chan-Zuckerberg Initiative, NIH, and the White House \cite{kaggle9}.

There have been projects such as a work at John Hopkins University that are focused on providing US county-level summaries of COVID-19 pandemic information and important attributes \cite{killeen2020county,wang2020comparing}.

The information in social networks has also been used in predicting the number or COVID-19 cases in mainland China \cite{shen2020using}. The work in \cite{pourhomayoun2020predicting} is also focused on an AI-based approach for predicting mortality risk in COVID-19 patients. 

There have been numerous approaches to model the pandemic using AI that have the historical outbreak information at the core of their analyses, such as the modified versions of SEIR model
and ARIMA-based analysis \cite{wang2020comparing,wang2020spatiotemporal,covid19simulator,srivastava2020learning,srivastava2020fast,kufel2020arima}. 
This work is distinguished from the mentioned projects and the majority of statistical works in this area in the sense that it is targeting the role of region-based features in the Spatio-temporal analysis of the pandemic with minimal use of historical data on the outbreak events. The area unit of this work is US county which enables a more fine-grained prediction scheme compared to the other works that have mostly targeted the state-level analytics. To our best knowledge, the works in \cite{wang2020spatiotemporal} and \cite{pei2020initial} are the only attempts in county-level modeling of the disease dynamics. In \cite{wang2020spatiotemporal}, authors have proposed a non-parametric model for epidemic data that incorporates area-level characteristics in the SIR model. The work in \cite{pei2020initial} uses a combination of iterated filtering and the Ensemble Adjustment Kalman filter for tuning their model, and their approach is based on a county-level SEIR model. The empirical results show that our approach outperforms these models on the evaluation benchmarks while providing a framework for utilizing deep learning in analysis and modeling the short-term pandemic events. We have made our codes and data publicly available and regularly maintained to help to expedite the research in this area.

\section{Methodology}

\subsection{Data}
This study focuses on analyzing the regions of the United States with statistical and AI-based approaches to obtain results and representations associated with their pandemic-related behavior. A primary and essential step in doing so is to prepare a dataset covering a wide range of information topics, from socio-economic to regional mobility reports.
More details regarding the primary data sources from which we have obtained information for this work's dataset are elaborated upon hereunder.

\subsubsection{COVID-19 Daily Information per County}
Our first step towards the mentioned objective is to gather the daily COVID-19 outbreak data. This data should include the number of cases that are confirmed to be caused by the novel coronavirus and its associated death toll. We are using the publicly accessible dataset API in \cite{covidapi,yang2020covidnet} to fetch the relevant data records. The table of data obtained using this API contains the numerical information along with dates corresponding to each record, and each document includes the number of confirmed cases and the number of deaths that occurred due to COVID-19 on that date. It also includes the number of recoveries from COVID-19 in the same format.
This dataset's significance is that it provides us with a detailed and high-resolution temporal trajectory of the COVID-19 outbreak in different urban regions across the United States. Using the dates, one can constitute a set of time-series for every county and monitor the outbreak along with the other metadata to make relevant inferences.

\subsubsection{US Census Demographic Data}
The US Census Demographic Data gathered by the US Census Bureau \cite{kaggle8} plays a critical role in our analysis by providing us with necessary information on each region's population. Additionally, this information includes specific features such as the types of work people in that region mainly take part in, their income levels, and other invaluable demographical and social information.

\subsubsection{US County-level Mortality}
The fluctuations in the mortality rate of a region is also a potential critical feature in pandemic analytics. The US county-level mortality dataset was incorporated into our collection to add the high-resolution mortality rate time-series throughout the years \cite{kaggle6,kaggle7}.
The age-standardized mortality rates provide us with information on variables, the values of which can be considered as the effects of specific causes. It is crucial since some of these causes might have contributed to the faster spread of COVID-19 in different regions \cite{dwyer2016us}.

\subsubsection{US County-Level Diversity Index}
Another dataset that offers a race-based breakdown of the county populations is available at \cite{kaggle4} with the diversity index values corresponding to the notion of ecological entropy. For a particular region, if K races comprise its population, the value of diversity index can be computed using the following formula:
$$ d_i = 1 - \sum_{i=1}^K (\cfrac{n_i}{N})^2 $$
In the above formula, $N$ is the total population and $n_i$ is the number of people from race $i$. This formula represents the probability $p$, which means that if we randomly pick two persons from this cohort, they are of different races with probability $p$.
In addition to that, we have the percentages of different races in the regional population as well.

\subsubsection{US Droughts by County}
Another source of valuable information regarding the land area and water resources per county is the data gathered by the US drought monitor \cite{kaggle432,kaggle3}. This data is incorporated into our collection as well.

\subsubsection{Election}
Based on the 2016 US Presidential Election, a breakdown of county populations' tendencies to vote for the main political parties is available \cite{electiondata}. These records are added to our collection as the democratic-republican breakdown of regional voters can reflect socio-economic and demographical features that form the underlying reasons for the regional voting tendencies.

\subsubsection{ICU Beds}
Since COVID-19 imposes significant problems in terms of the extensive use of ICU beds and medical resources such as mechanical ventilators, having access to the number of ICU beds in each county is helpful. This information offers a glance at the medical care capacity of each region and its potential to provide care for the patients in ICUs \cite{kaggle2}. It could be argued that having knowledge of the ICU-related capacity of regional healthcare providers can, to some extent, represent the amount of their COVID-19 related resources, such as ventilators and other needed resources.

\subsubsection{US Household Income Statistics}
The aggregate dataset on central statistical values on the US household income per county (including average, median, and standard deviation) is used to provide information on the financial well-being of the affected regions' occupants \cite{kaggle1}.

\subsubsection{COVID-19 Hospitalizations and Influenza Activity Level}
Aside from the socio-economical and demographical features of a region, the number of active and potential COVID-19 cases is a critical factor. This information can be leveraged to provide a possible threat level for the region. These records are made available by CDC for specific areas and are incorporated into our collection as well \cite{cdccovidushospitalization,cdccovidushospitalizationlabconfirmed}.

\subsubsection{Google Mobility Reports}
The COVID-19 virus is highly contagious. Therefore, the self-quarantine and social distancing measures are principal effective methodologies in bolstering the prevention efforts. Our collection includes Google's mobility reports obtained from \cite{killeen2020county}. These records elaborate on the mobility levels across US regions, which are broken down into the following categories of mobility:
\begin{enumerate}
    \item Retail and Recreation
    \item Grocery and Pharmacy
    \item Parks
    \item Transit Stations
    \item Workplaces
    \item Residential
\end{enumerate}

In addition, we have computed a compliance measure that has to do with the overall compliance with the shelter at home criteria:

$$ \text{compliance} = -1 - \cfrac{(1/6)\sum_{i=1}^6m_i - 100}{100.0}$$

In the above formula, $m_i$  is the mobility report for the $i$th mobility category. This value is computed through time to provide an overall measure of mobility through time. The compliance measures of $+1$ and $-1$ mean $+100\%$ and $-100\%$ changes from the baseline mobility behavior, respectively.

\subsubsection{Food Businesses}
Restaurants and food businesses are affected severely by the economic impacts of this outbreak. At the same time, they have not ceased to provide services that are essential and required by many. To reach a proper perspective of the food business in each region, we have prepared another dataset based on records in \cite{nsrestaurant} to provide statistics on regional restaurant revenue and employment. Analysis of restaurants’ status is important in the sense that they are mostly public places that host large gatherings, and in the time of a pandemic, their role is critical.

\subsubsection{Physical Activity and Life Expectancy}
Various features have been selected from the dataset in \cite{healthdata} to reflect on the obesity and physical activity representation for different US regions. These features include the last prevalence survey and the changes in patterns. Also, Life Expectancy related features are valuable information for representing each region. They are included as well in our analyses.

\subsubsection{Diabetes}
Different features to represent a region according to the diabetes-related characteristics were selected from the data in \cite{healthdata}. These include age-standardized features and clusters that have to do with diabetes-related diagnoses. 

\subsubsection{Drinking Habits}
Information on regional drinking habits from 2005-2012 has also been used in this work \cite{healthdata}. This information includes the proportions of different categories of drinkers clustered by sex and age. The categories are as follows:
\begin{itemize}
    \item “Any”: a minimum of one drink of any alcoholic beverage per $30$ days
    \item “Heavy”: a minimum average of one drink per day for women and two drinks for men per $30$ days
    \item "Binge”: a minimum of four drinks for women and five drinks for men on a single occasion at least once per $30$ days
\end{itemize}

\begin{table}[]
    \centering
    \caption{Overview of the Features}
    \label{tab:overviewfeatures}
    \begin{tabular}{l|l}
\textbf{Category}                                   & \textbf{Description  }                                                                                            \\ \hline
\multirow{3}{*}{Food Businesses (static)}  & Food   and Beverage Locations                                                                            \\
                                           & Restaurant   Employments                                                                                 \\
                                           & Sale   and Economy                                                                                       \\ 
Gender (static)                            & Percentage   of Male and Female                                                                          \\
Race (static)                              & Ratio   of different races                                                                               \\
Election (static)                          & Ratio   of Democratic, Republican, and other voters                                                      \\
\multirow{2}{*}{Income (static)}           & Wage   Statistics                                                                                        \\
                                           & Poverty   Information                                                                                    \\
Commute (static)                           & Statistics   of Methods of Commute to Work and Their Ratio                                               \\
Hospitals and Mortality (static)           & Information on ICU Capacity and Statistics   on Region’s Mortality                                       \\
Obesity and Physical Activity (static)     & Information on the Statistics of Obesity   and Physical Activity and the Changes in Patterns             \\
Life Expectancy (static)                   & Regional Life Expectancy Values in Years                                                                 \\
Drinking (static)                          & Alcohol Consumption Patterns and Changes                                                                 \\
Diabetes (static)                          & Patterns of Different Types of Diabetes   Diagnoses and Changes in Them                                  \\
Land and Water (static)                    & Information on Land and Water Resources of   Regions                                                     \\
Employment (static)                        & Ratio of Different Job Types and Other   Statistics                                                      \\
CDC Hospitalizations and Surveys (dynamic) & Number of Hospitalizations due to COVID-19   and Influenza Activity Surveys                              \\
Google Mobility Reports (dynamic)          & Breakdown of Regional Mobility in Different   Categories Based on Which Our Compliance Score Is Computed
\end{tabular}
    
\end{table}

\subsubsection{Analytics}
In what follows, the analytical techniques that we have designed and used in this work are explained. To draw meaning from the data that we have at hand, we have designed and utilized a variety of techniques. These methodologies range from traditional statistical methodologies to the design and testing of deep learning inference pipelines for event prediction. 
We select a set of representative features to use in our analytics from the gathered collection of datasets. More details on the nature of these features are shown in Table \ref{tab:overviewfeatures}.

\subsubsection{Feature Informativeness for Sub-region Representation}
An important question that is raised in analyzing a dataset with well-defined categories of features is how important these features are in describing the entities associated with them. From the particular perspective of enabling the differentiation between two regions, it can be said that a measure of importance is the contribution of each one of these selected features to the overall variation in datapoints. The boundary case is that if a feature always has the same value, it is not informative as there is no entropy value associated with its distribution.
To begin with, we associate a mathematical vector with each data point, which contains the values of all its dynamic and static features associated with a specific date and location. Since we are mainly targeting US counties in this study, each record would be associated with a US county at a specific date. We then use Linear Principal Component Analysis \cite{wold1987principal} to reduce the dimensionality of these data points and to evaluate the importance of the selected features in terms of their contribution to the overall variation.
Results show that in order to retain over $98\%$ of the original variance, a minimum of $55$ principal components should be considered. Each one of these components is found as a linear combination of the original set of features, and that along with the percentage of variance along the axis of that component can be used as a measure of performance.
To be more specific, considering $n$ features and m data points that result in $p$ PCA components to retain $98\%$ of the variation, we will have:
$$
\vec{c_i} = \langle v_1,v_2,\cdots,v_n \rangle \in \mathbb{R}^n
$$

And $u_i$ is the total variance along the axis of $i$th PCA component. This can be thought of as a measure of importance for the PCA components, and the absolute value of $v_i$s magnitudes can be considered as the importance of original feature $i$'s contribution to its making. Therefore, we will have the following measure of informativeness defined for our features:

$$ \vec{I} \in \mathbb{R}^n $$
$$ \vec{I} = \sum_{j=1}^p u_j \cdot \vec{c_j} $$
The features can be sorted according to these values, and the categories can also be considered in their relevant importance.
Note that this is just one definition of informativeness; for example, certain features might not vary a lot, but when they do, they are potentially associated with severe changes in the COVID-19 events. Therefore, the importance score that has been captured here merely has to do with how better we are able to distinguish between locations based on a feature.

\subsection{Statistical Analytics}
In order to better understand the co-occurrences of the features in our input dataset and their corresponding COVID-19 related events, we have performed an in-depth correlation analysis on them. We have considered four principal measures of correlation, namely: Pearson, Kendall, Histogram Intersection, and Spearman, as described in Table \ref{tab:correlations}. We have used the Pearson correlation coefficient along with the p-values to shed light on the presence or absence of a significant relationship between the values of each specific feature and each category of pandemic outcome. We have also computed nonparametric Spearman rank correlation coefficients between any two of our random variables. This value would be computed as the Pearson measure of the raw values converted to their ranks. The formulation is shown in Table \ref{tab:correlations} in which $d_i$  is the difference in paired ranks. Mutual information has also been used to provide additional information on such relationships.
This coefficient measures the strength of the association between the values of these random variables in terms of their ranks. Since many of the relationships in our dataset can be intuitively thought of as monotonic, these values are particularly important. To better understand the concordance and discordance, Kendall correlation is computed as well. In the formulation shown in Table \ref{tab:correlations}, $m_1$ and $m_2$ are the numbers of concordant and discordant pairs of values, respectively. Normalized Histogram Intersection is another methodology directly targeting the distributions of these variables. The degree of their overlap represents how closely $x$’s distribution follows the distribution of $y$. It has also been utilized in finding the results of this section.

\begin{table}[]
    \centering
    \caption{The equations for the three main correlation analysis techniques used in this work, namely, Pearson, Spearman, and Kendall correlations to evaluate the monotonic and general relationships between variables.}
    \label{tab:correlations}
    \begin{tabular}{c|c}
        \textbf{Correlation Analysis} & \textbf{Formula} \\ \hline
        Pearson & $r_{x,y}=\cfrac{
            \sum_{i=1}^m (x_i - \mu_x)\cdot(y_i-\mu_y)
        }{
        \sqrt{
        
        }(\sum_{i=1}^m (x_i - \mu_x)^2) \cdot (\sum_{i=1}^m (y_i - \mu_y) ^ 2)
        }$ \\ 
        
        Spearman & $s_{x,y} = 1 - \cfrac{6 \sum d_i^2}{m(m^2-1)}$ \\
        Kendall & $k_{x,y}=\cfrac{m_1-m_2}{\binom{m}{2}}$
    \end{tabular}
    
\end{table}

\subsection{Neural Event Prediction}
In continuation of our statistical analyses on COVID-19 event distributions, we have designed a neural inference pipeline to help with the effective utilization of both learned deep representations and the embedded sequential information in the dataset.

In this work, we introduce a neural architecture, which is trained and used for COVID-19 event prediction across the US regions. The Double Window Long Short Term Memory COVID-19 Predictor (DWLSTM-CP) is comprised of multiple components for domain mapping and deep processing. First, using its dynamic projection which is a fully connected layer, the dynamic feature vectors which reflect on temporal dynamics will be mapped to a new space and represented with a further concise mathematical vector.

$$ \langle \tilde{x}^{\text{dynamic}}_1, \tilde{x}^{\text{dynamic}}_2,\cdots,\tilde{x}^{\text{dynamic}}_T \rangle = F^{\text{dynamic}}_{\text{projection}}(
    \langle x^{\text{dynamic}}_1, x^{\text{dynamic}}_2,\cdots,x^{\text{dynamic}}_T \rangle
)$$

This step is essential due to the fact that an optimal deep inference pipeline is the one that retains only the information required by each level and minimizes redundancies \cite{tishby2015deep}. The projections are designed to help the network achieve this objective. These are then fed to the LSTM core for processing.

$$ A^{\text{dynamic}} = \text{LSTM}(\tilde{X}) $$

\begin{figure}
    \centering
    \includegraphics[width=0.8\textwidth]{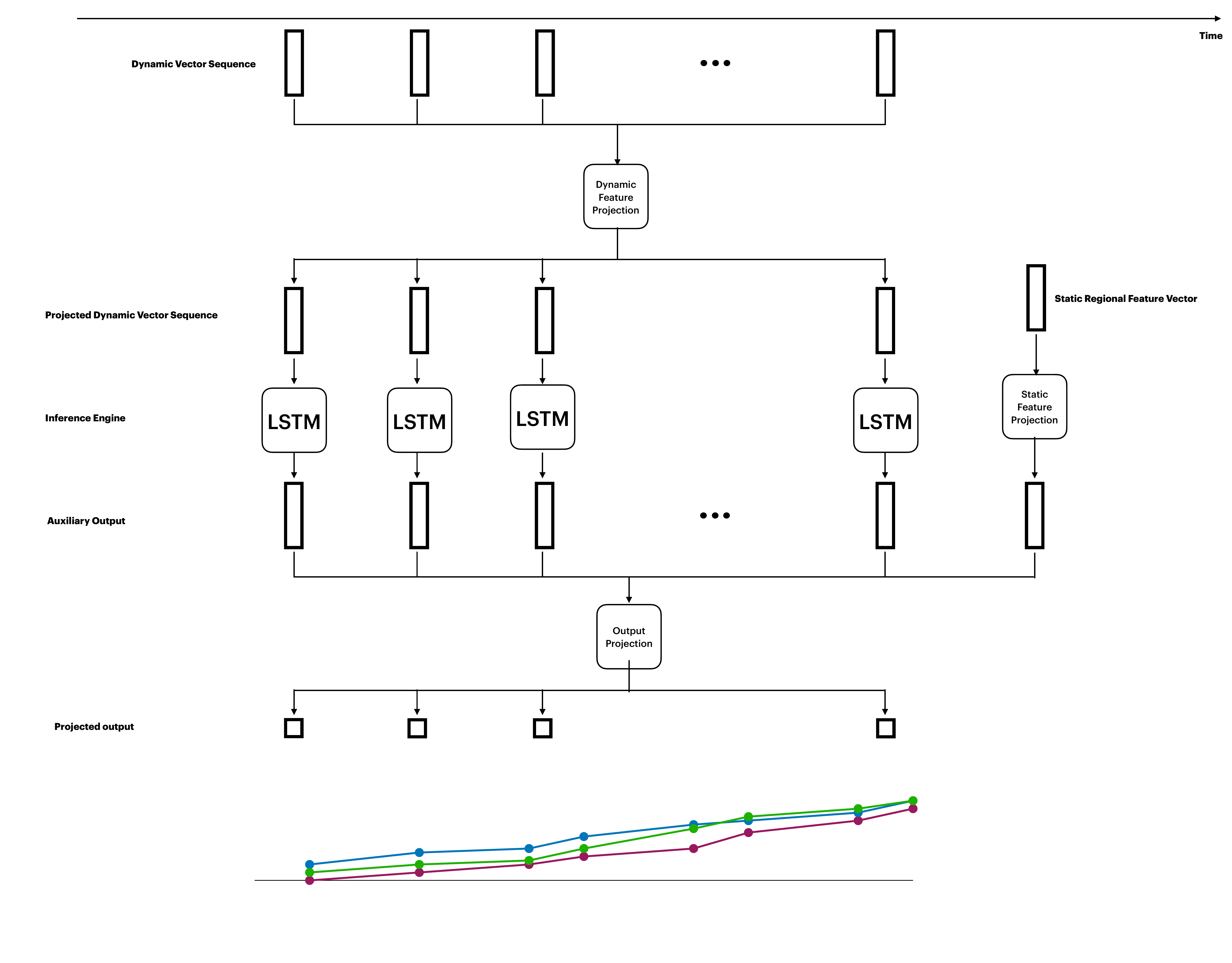}
    \caption{The full inference pipeline of our Double Window LSTM-based COVID-19 event prediction is shown in this block diagram.}
    \label{fig:pipeline}
\end{figure}

Each one of these outputs is concatenated with the projected version of static features, $F_{\text{projection}}^{\text{static}}(x^{\text{static}})$, and fed to the output regression unit. The outputs are compared with the ground truth time-series, and a weighted Mean Squared Error loss along with Norm-based regularization is used to guide the training process while encouraging more focus on the points with large values. The overall pipeline is shown in Figure \ref{fig:pipeline}.

It is worth mentioning that this approach leverages and utilizes all of the features discussed in the previous sections. It learns representations that take various factors, from different categories of mobility and activities to socio-economic information, to make accurate short-term predictions while reducing the need for lengthy historical data on the pandemic outcomes. There are many occasions in which accurate and reliable historical data on the pandemic is not available due to a variety of reasons (e.g., a problem in reporting scheme), which motivates approaches with less dependency on it.

\section{Results}
The results on our regional dataset in terms of feature importance from the principal component analysis indicate the following features contribute to the overall representation significantly:
\begin{itemize}
    \item Restaurant businesses, namely the contribution to the state economy and the count of food and beverage locations. Even though we only have access to state-level data, its importance can be intuitively argued as it reflects on the counties that the state includes. This is due to the fact that the status of restaurants plays an essential role in such pandemics. 
    \item The influenza activity level is another critical feature in the analysis. Given the similarity of symptoms between Influenza and COVID-19 infection, monitoring Influenza activity is very helpful for COVID-19 pandemic understanding.
    \item Diversity index, which signifies the probability of two randomly selected persons belonging to different races from a population, also plays a crucial role in representing the regions.
    \item The changes in the mortality rate that is not associated with COVID-19 are beneficial as well. This is also intuitively arguable as it can be thought of as a measure of mortality related sensitivity for the regions.
\end{itemize}

\begin{figure}
    \centering
    \includegraphics[width=0.7\textwidth]{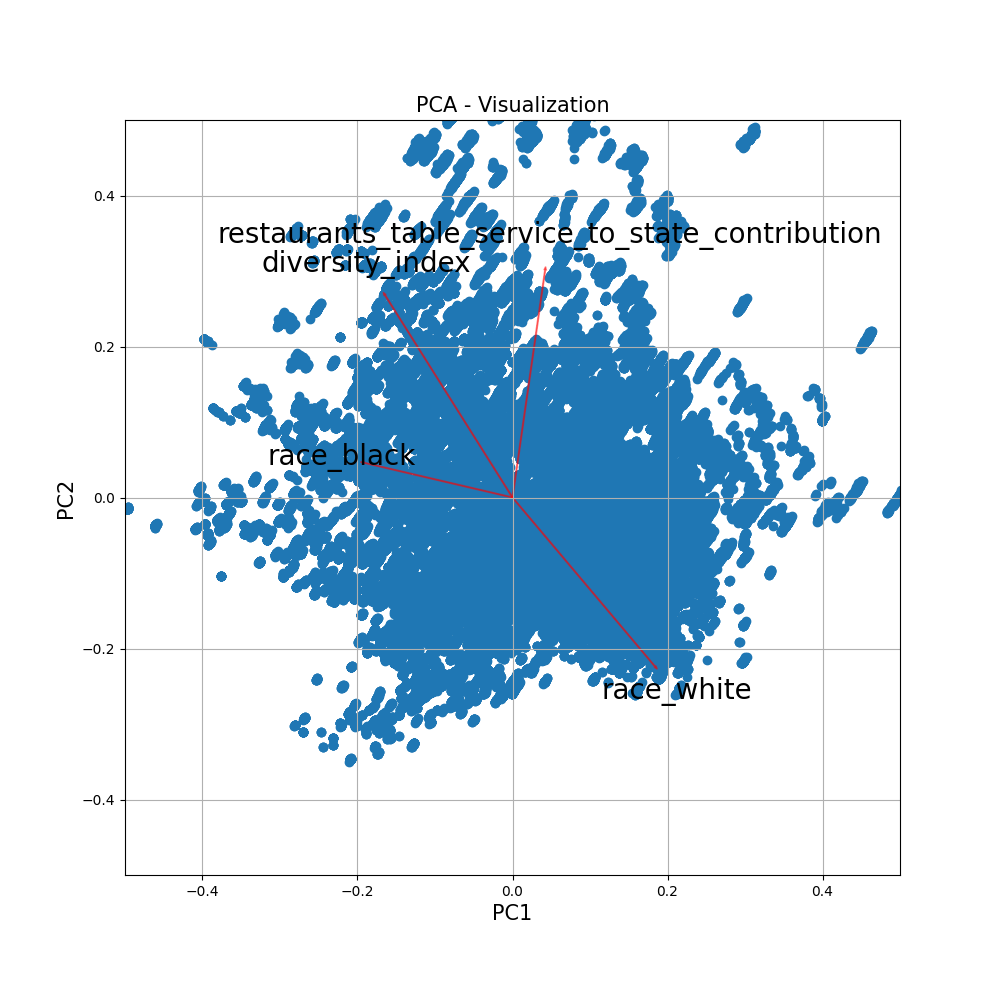}
    \caption{The plot in this figure is a PCA BiPlot which shows the variations of the first two PCA components and axes of some of the selected features.}
    \label{fig:pcavis}
\end{figure}

\begin{figure}
    \centering
    \includegraphics[width=0.5\textwidth]{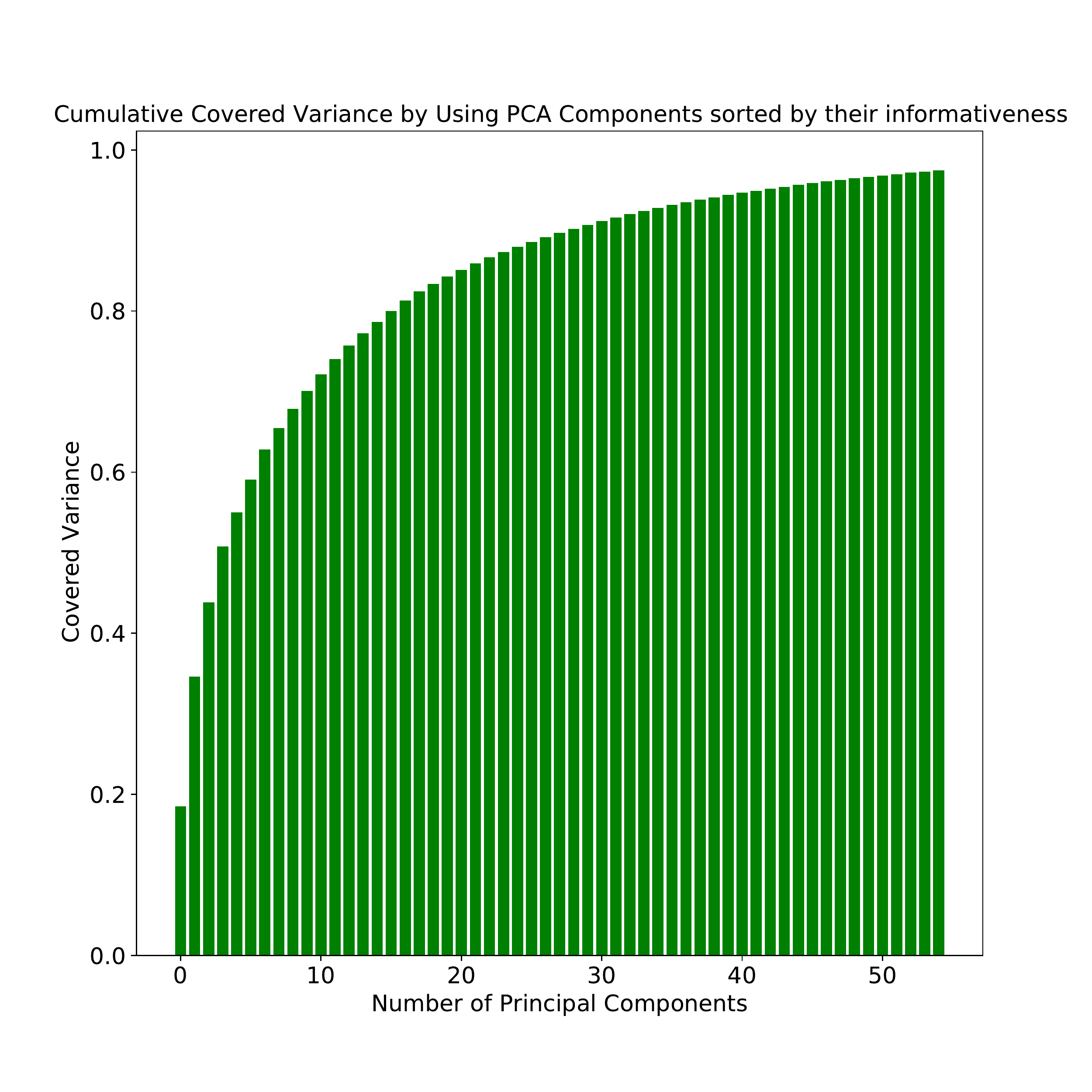}
    \caption{The cumulative amount of variance covered by using up to a certain number of PCA components. This is assuming that they are sorted by their corresponding eigenvalue, meaning that the first component contributes more to variance coverage than the ones selected after it.}
    \label{fig:componentstoretain}
\end{figure}

Figure \ref{fig:pcavis} shows how the projected points scatter after the PCA as well. The results indicate that $55$ PCA components are required to retain over $98\%$ of the variance of the dataset, and Figure \ref{fig:componentstoretain} shows the progress of covering the variance by adding each one of them sorted by their importance. Table \ref{tab:importancesamples} provides the values of the aforementioned importance metric computed for sample features in different importance levels.

\begin{table}[]
\centering
\caption{Sample Features of High and Low Informativeness Score}
\label{tab:importancesamples}
\begin{tabular}{lll}
\textbf{Level}                 & \textbf{Feature}                                                       & \textbf{Score} \\ \hline
\multirow{3}{*}{High} & Diversity Index                                               & 0.148 \\
                      & Contribution of   Restaurants’ Table Service to State Economy & 0.130 \\
                      & African American   Ratio                                      & 0.109 \\
\multirow{3}{*}{Low}  & Percentage of   Men                                           & 0.020 \\
                      & Pacific Islanders   Ratio                                     & 0.013 \\
                      & Percentage of   Family Jobs                                   & 0.006
\end{tabular}
\end{table}

\subsection{Statistical Analytics}
The results of correlation analyses help empirically and quantitatively validate many of the relationships mentioned in the known hypotheses regarding the COVID-19 outbreak. The Pearson correlation of $-28.67\%$ with the p-value of $0.046$ indicates a significant relationship between the percentage of food businesses in the state economy, and the average cumulative death count in its counties. Another example is the value of the Spearman correlation coefficients between the different types of commute to work associated with each county and the values of the pandemic-related events. From Table \ref{tab:statanal1}, it is apparent that there is a positive relationship between the proportion of public transit as a method of commute to work and the spread of COVID-19 in the region. Another example is the Pearson correlation between the ratio of different races in regions and the pandemic outcomes. It is known that COVID-19 is affecting the African American community disproportionately \cite{fourreasonsblacksaffected}. Accordingly, the values in Table \ref{tab:statanal2} show a higher correlation between the ratio of African Americans and the severity of COVID-19 outcomes.

\begin{table}[]
    \centering
    \caption{The Spearman correlation coefficients between the share of different methods of commute in county transportation and the cumulative pandemic outcomes. As we can see, the more the percentage of public transit is for the method of commute to work, the more the number of potential cases is expected to be as the Spearman correlation coefficient is an indicator of a monotonic relationship between variables.}
    \label{tab:statanal1}
    \begin{tabular}{c|c|c|c}
         & Cumulative Death Count & Cumulative Case Count & Cumulative Recovery Count \\ \hline
        Drive & $0.22$ & $0.20$ & $-0.03$ \\
        Carpool & $-0.04$ & $0.04$ & $0.04$ \\
        Transit & $0.20$ & $0.12$ & $-0.06$ \\
        Walk & $-0.29$ & $-0.35$ & $0.05$ \\
    \end{tabular}
    
\end{table}

\begin{table}[]
    \centering
    \caption{Pearson correlation between the race percentages per county and COVID-19 variables, which also indicates the more diverse regions were impacted the most. This result is in accordance with the findings of feature importance, which listed the Diversity Index as one of the most important entities.}
    \label{tab:statanal2}
    \begin{tabular}{c|c|c|c}
          & Cumulative Death Count & Cumulative Case Count & Cumulative Recovery Count \\ \hline
         White & $-0.30$ & $-0.48$ & $-0.15$ \\
         African-American & $0.34$ & $0.42$ & $-0.01$ \\
         Hispanic & $0.04$ & $0.23$ & $0.2$ \\
         Native American & $0.03$ & $0.02$ & $0.03$ \\
         Asian & $0.14$ & $0.11$ & $0.04$ \\
         Pacific Islander & $-0.03$ & $-0.02$ & $0.03$ \\
    \end{tabular}
    
\end{table}

\subsection{Neural Event Prediction}
\begin{figure}
    \centering
    \includegraphics[width=1\textwidth]{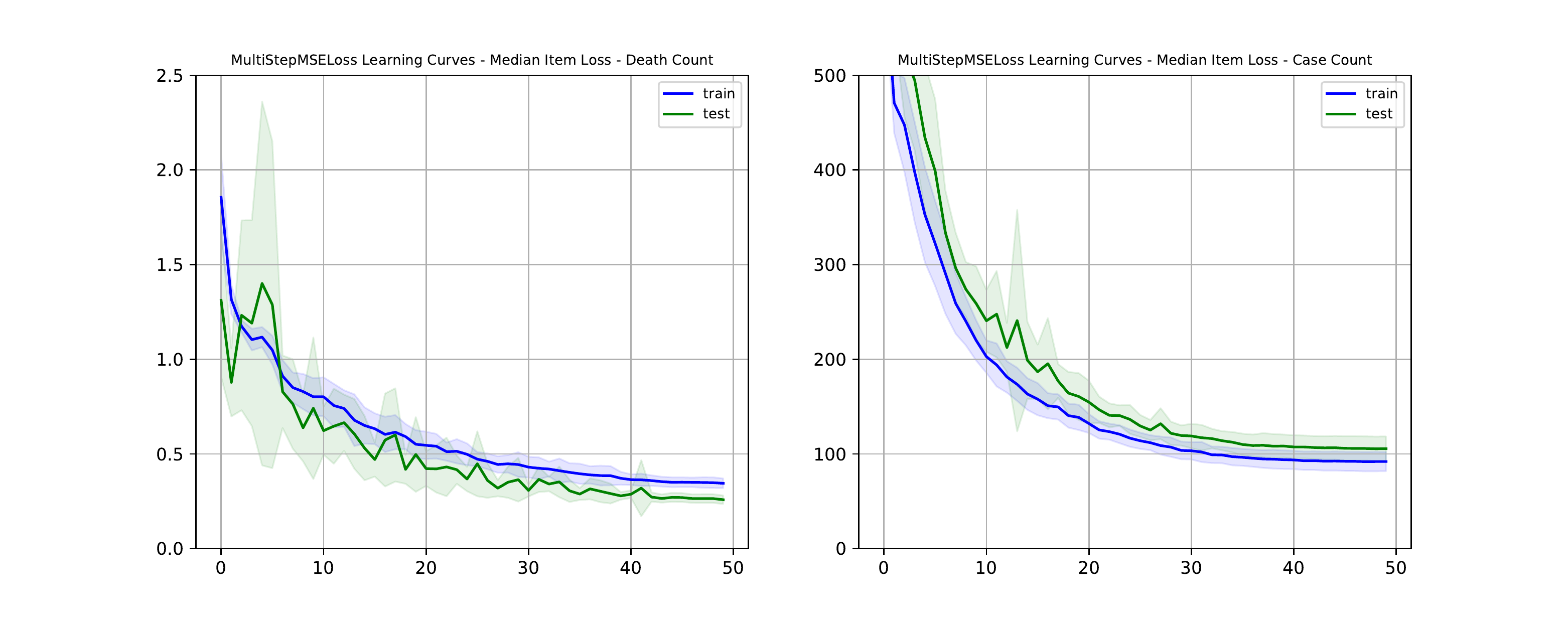}
    \caption{Sample Test Prediction of Cumulative Death Count per 100k Population - Four regions exhibiting different severity levels are chosen to show the efficacy of the model. The $95\%$ confidence intervals for ARIMA* and DWLSTM models are shown and clearly indicate the stability in training our model and the predictions made by it.}
    \label{fig:learningcurves}
\end{figure}

The collected set of datasets in this work provide a sufficient number of records for enabling the efficient use of Artificial Intelligence for Spatio-temporal representation learning. We show this by training instances of our proposed DoubleWindowLSTM architecture on the two main short-term tasks regarding epidemic modeling; namely, new daily death and case count. In our dataset, we considered the US COVID-19 information from March 1st, 2020 to July 22nd, 2020, in which the July data is used for our evaluations, and the rest are leveraged for training and cross-validation. The objective using which the proposed architecture was trained is a multi-step weighted Mean Squared Error (MSE) loss, which helps to minimize a notion of distance between the predictions and the target ground-truth while encouraging (by assigning larger weights) to the windows that exhibit larger values. These thresholds are empirically tuned and set prior to the training procedure. The learning curves for both experiments indicate clear convergence in Figure \ref{fig:learningcurves}. 

To quantitatively evaluate the performance, we have reported the Root Mean Square Error (RMSE) for the prediction of new daily deaths and cases due to COVID-19 in Table \ref{tab:arimadwlstm}. For comparison, we have used the ARIMA model as well with the parameters set according to the work in \cite{kufel2020arima} that have fine-tuned this scheme for forecasting the dynamics of COVID-19 cases in Europe. We have also found the best ARIMA model in each scenario according to Augmented Dickey-Fuller (ADF) tests and based on Akaike information criterion (AIC) and reported the results denoted by ARIMA*. To compare with other works in this area, we had to aggregate our county-level findings to form estimators for state-level prediction. From the results reported in Table \ref{tab:dwlstmstate}, it is interesting to observe that the aggregated estimator based on our model achieves strong evaluation result comparable to the models that achieve highest scores, while clearly outperforming the other two models that are inherently county-level, namely, the works in \cite{wang2020spatiotemporal} and \cite{pei2020initial}.

\begin{figure}
    \centering
    \includegraphics[width=1\textwidth]{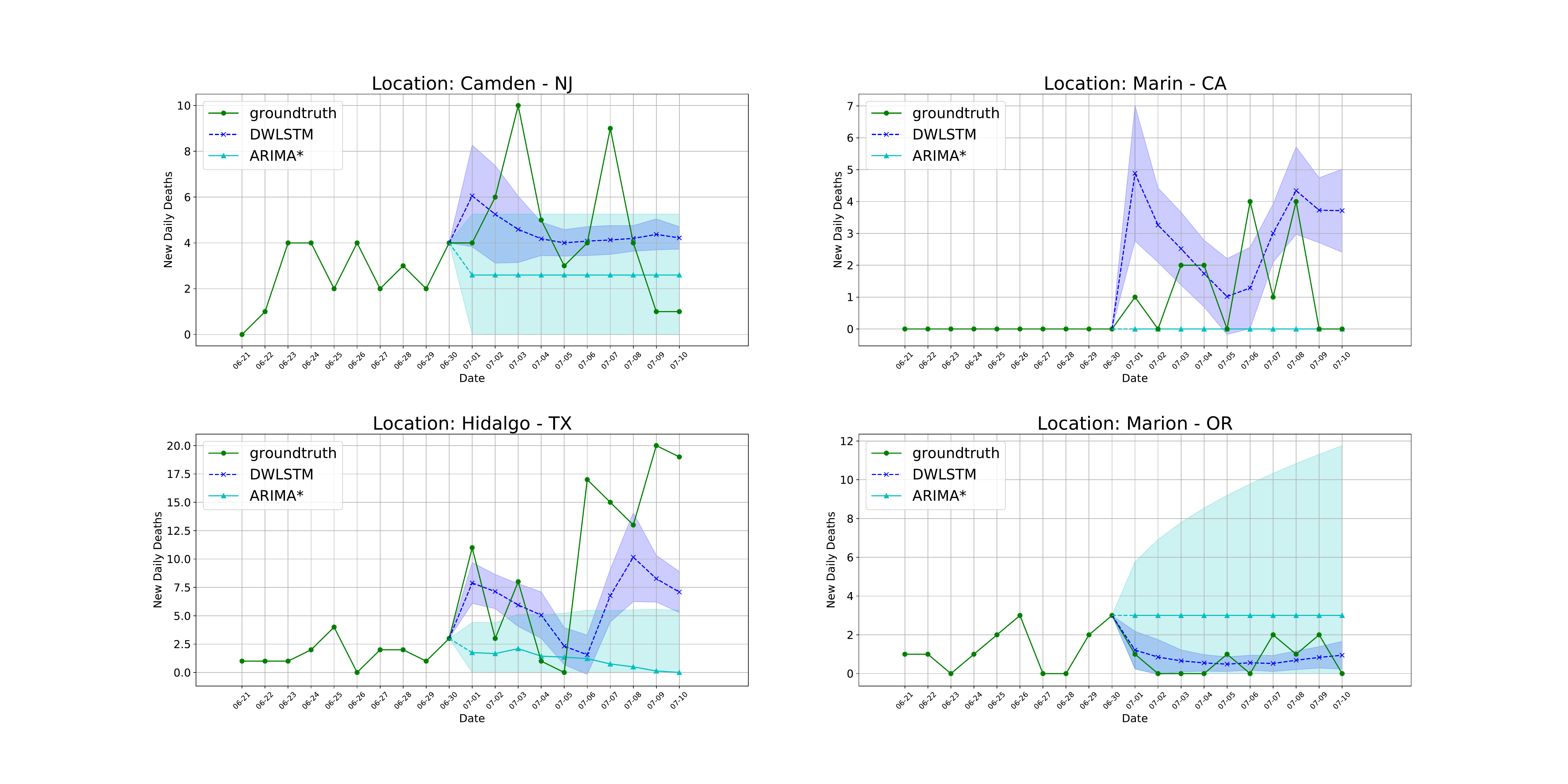}
    \caption{Sample Test Prediction of Cumulative Death Count per $100$k Population - Four regions exhibiting different severity levels are chosen to show the efficacy of the model. The $95\%$ confidence intervals for ARIMA* and DWLSTM models are shown and clearly indicate the stability in training our model and the predictions made by it.}
    \label{fig:samplepreds}
\end{figure}

The predictions for severak regions exhibiting different severities are shown in Figure \ref{fig:samplepreds}. These results can help the reader in a qualitative assessment of the model performances, in which the outputs of our approach demonstrate high stability and follow the trajectory of the ground-truth with precision.

\begin{table}[]
    \centering
    \caption{The comparison of Average Daily Root Mean Square Error between the DWLSTM predictions and the ARIMA-based predictions. The evaluation is performed on the test set, which includes the data from the end of June 2020 to July 22nd, 2020.}
    \label{tab:arimadwlstm}
    \begin{tabular}{lllll}
\multirow{2}{*}{\textbf{Objective and Timeframe}} & \multicolumn{2}{l}{\textbf{New Daily Deaths}} & \multicolumn{2}{l}{\textbf{New Daily Cases}} \\
                                         & 10-day            & 15-day           & 10-day           & 15-day           \\ \hline
DWLSTM                                   & 4.4347            & 3.0435           & 81.4205          & 92.4027          \\
ARIMA*                                   & 29.9813           & 12.7631          & 233.3008         & 235.3828         \\
ARIMA(1,2,0)                             & 57.1886           & 22.4285          & 394.3747         & 566.5686        
\end{tabular}
    
\end{table}

\begin{table}[]
    \centering
    \caption{An overview of the comparisons for the evaluation results on DWLSTM compared to eight COVID-19 prediction models is shown in this table. The evaluation for the DWLSTM has been done for the next 15 days during the month of July until July 22nd, in which many drastic changes to the pattern of the outbreak have been observed in the US, especially in California and Texas. The other models are evaluated until June 28th and on different datasets on pandemic events, namely, Johns Hopkins University (JHU) \cite{yang2020covidnet,covidapi}, New York Times dataset (NYU) \cite{nytimesdataset}, and the US Facts dataset (USF) \cite{usfacts}. It should also be noted that even though the objective for the DWLSTM model was to predict county-level information, the provided state-level errors which are obtained by aggregation fall in the range of the dominant COVID-19 predictor models that rely heavily on the accuracy of the historical epidemic data.}
    \label{tab:dwlstmstate}
    \begin{tabular}{llll}
         & \textbf{Prediction Window (days)} & \textbf{Average Daily RMSE} & \textbf{Ground-truth Source}\\ \hline
        DWLSTM & 15 & 26.23 & JHU \\
        SIkJa & $14$ & $23.63$ & JHU \\
        UCLA SuEIR & $14$ & $22.97$ & NYT \\
        CovidActNow SEIR CAN & $14$ & $27.78$ & NYT \\
        IowaStateLW. STEM & $14$ & $26.67$ & JHU \\
        Covid19Sim Simulator & $14$ & $27.82$ & JHU \\
        JHU IDD CovidSP & $14$ & $48.97$ & USF \\
        CU Select & $14$ & $32.36$ & USF
    \end{tabular}
    
\end{table}

\section{Discussion}
\subsection{Principal Findings}
The primary objective of this work is focused on leveraging regional representations for accurate short-term predictive modeling of the epidemic with minimal use of historical data. It is plausible to assume that the features chosen in this work, which reflect on different characteristics of a region, include valuable information for efficient prediction of pandemic events. The static features include various socio-economic and demographical properties associated with a region and its population. Combined with the dynamic set of features such as influenza activity level and mobility patterns, this information was leveraged along with a short track of pandemic time-series for predictive modeling. We do not claim that the data points coming from this domain are statistically sufficient for the pandemic event prediction tasks; however, empirical results indicate that they can be effectively utilized for these objectives. There are occurrences outside of this domain that can impact the outcomes (e.g., the initial impact of a large number of infected people arriving in a specific location is not initially captured by our scheme). Nevertheless, the results indicate that the data points coming solely from this work's domain can help in the effective knowledge extraction regarding the current and future values of pandemic-related time-series. The result section elaborated on the statistical findings and introduced a measure of feature importance. In addition, a neural network architecture that has a long short-term memory configured recurrent neural network in its core was introduced to serve as a new baseline for COVID-19 event prediction.

\subsection{Comparison with Previous Studies}
Since the beginning of the COVID-19 outbreak, there have been works focusing on gathering information or performing statistical analysis related to this epidemic. This work is focused on learning and analysis of the high-resolution spatiotemporal representation of urban areas. We provide a collection of datasets and select a large number of features to reflect on various demographics, socio-economics, mobility, and pandemic information. We have used statistical analysis techniques to investigate the relationships between individual features and the epidemic, while also considering the contribution of such features to the overall representation power. We have also proposed a deep learning framework to validate this idea that such region-based representations can be leveraged to obtain accurate predictions of the epidemic trajectories while using but a minimal amount of historical data on the outbreak events (e.g., number of cases). Even though are model is trained with the objective of providing county-level predictions, we have aggregated these county-level predictions and used these now state-level estimators to evaluate the loss on the most recent data. In Table 6, we have compared these results with the information on the similar performance measure of the eight COVID-19 prediction works that perform state-level inference making. It can be seen that our framework provides a simple solution which outperforms the other county-level methodologies (namely, \cite{wang2020spatiotemporal} and \cite{pei2020initial}) on this task.

\subsection{Applications}
The importance of clearly defined policies enforced at the proper time on alleviating the adverse impacts of a pandemic in different areas is crystal clear. One of the important applications of this work is in providing researchers and agencies with a more in-depth understanding of the co-occurrence of idiosyncratic patterns associated with regions and the predicted pattern of the outbreak. This information can be used to assist policymakers, for example, to render the details of their decisions such as lockdowns, more fine-grained and attuned to the regional needs. These include the intensity and length of enforcing such measures. The ability to predict pandemic-related occurrences (e.g., number of deaths, cases, and recoveries) is another valuable application of this work. This knowledge will provide hospitals and healthcare facilities with targeted information to help with the efficient allocation of their resources. Another important application of this work is when there is a lack of availability for  accurate and reliable historical data on the epidemic events. For example, when it is realized that the previous reports on the number of cases and deaths due to the pandemic were not reliable, such finding will not affect our solution due to its less degree of dependence on the historical data on the epidemic than other models which base their analysis on them at the core of their analyses.

\subsection{Limitations}
This study has several limitations that should be discussed. The initial notion of feature informativeness which was discussed in the earlier sections of this article mainly has to do with the contribution of features to the variance in representing regions and areas. Given the nature of this study, combining this and the relationship between them and the pandemic and providing more in-depth prior domain knowledge can help with a better definition of feature importance. Our methodology provides a means to use  region-based representations to obtain predictions with less reliance on the historical epidemic data. Nevertheless, generalizing the network architecture in this work and providing access to more extended and reliable historical data, if possible, can be an improvement and is worthwhile as a potential future direction. Utilizing attention-based methodologies and other interpretation techniques with the pre-trained weights is also a well-suited future direction to better understand what the models learn.

\subsection{Conclusions}
In this study, we gathered a collection of datasets on a wide range of features associated with US regions. Our approach then used various statistical techniques and machine learning to measure the relationship between these regional representations and the pandemic time-series events and perform predictive modeling with minimal use of historical data on the epidemic. Both quantitative and qualitative evaluations were used in assessing the efficacy of our design, which renders it suitable for applications in various areas related to pandemic understanding and control. This is crucial since the information on the patterns and predictions related to an outbreak play a critical role in elaborate preparations for the pandemic, such as improving the allocation of resources in healthcare systems that will otherwise be overwhelmed by an unexpected number of cases.

\appendices
\section{Early-phase Analytics}
It is important for a predictive modeling approach on the pandemics to be able to help when the epidemic is in its early stages. To evaluate the performance of our approach, we have performed experiments on the early stages of the COVID-19 pandemic as well. In this particular dataset, the March 1st, 2020 to May 5th, 2020 date range is covered. Using a k-fold validation approach, the performance of the model is evaluated and reported in Table \ref{tab:earlyperformance}. It is shown that the network operates significantly better than ARIMA*, the details of which were discussed in the article. Please note that ARIMA based models have shown success in predicting COVID-19 events in the literature.

\begin{table}[h]
    \centering
    \caption{This table shows the average Daily Root Mean Square Error for the DWLSTM model compared to the ARIMA* predictions. The evaluations are done using a dataset that contains only the early stages of the COVID-19 outbreak in the US. The objective in the following experiments was to predict the new daily death counts for the US counties.}
    \label{tab:earlyperformancedeath}
\begin{tabular}{llclc}
\multirow{2}{*}{} & \multicolumn{2}{l}{\textbf{DWLSTM}}                      & \multicolumn{2}{l}{\textbf{ARIMA*}}                       \\
                  & \textbf{Macro}              & \multicolumn{1}{l}{\textbf{Micro}} & \textbf{Macro}               & \multicolumn{1}{l}{\textbf{Micro}} \\ \hline
\textbf{$10$-day window}      & \multicolumn{1}{c}{$15.62$} & $38.12$                             & \multicolumn{1}{c}{$91.06$}    & $237.07$                          \\
\textbf{$15$-day window}      & \multicolumn{1}{c}{$16.80$} & $40.72$                           & \multicolumn{1}{c}{$120.92$} & $339.51$                          
\end{tabular}
    
\end{table}

\begin{table}[h]
    \centering
    \caption{This table shows the average Daily Root Mean Square Error for the DWLSTM model compared to the ARIMA* predictions. The evaluations are done using a dataset that contains only the early stages of the COVID-19 outbreak in the US. The objective in the following experiments was to predict the new daily confirmed COVID-19 case counts for the US counties.}
    \label{tab:earlyperformancecases}
\begin{tabular}{llclc}
\multirow{2}{*}{} & \multicolumn{2}{l}{\textbf{DWLSTM}}                      & \multicolumn{2}{l}{\textbf{ARIMA*}}                       \\
                  & \textbf{Macro}              & \multicolumn{1}{l}{\textbf{Micro}} & \textbf{Macro}               & \multicolumn{1}{l}{\textbf{Micro}} \\ \hline
\textbf{$10$-day window}      & \multicolumn{1}{c}{$70.44$} & $107.34$                           & \multicolumn{1}{c}{$184.27$} & $271.87$                           \\
\textbf{$15$-day window}      & \multicolumn{1}{c}{$91.45$} & $134.22$                           & \multicolumn{1}{c}{$512.09$} & $1215.06$                          
\end{tabular}
    
\end{table}

\section{Different Pandemic Events}
In the first appendix, the performance of the model on the two main tasks regarding COVID-19 predictions and simulations was demonstrated. To add on that, Table \ref{tab:cumsumpred} shows the performance of the model on the task of predicting normalized cumulative death counts for each county which is attributed to the pandemic. The other factor that is shown in Table \ref{tab:cumsumpred} is the variations of the performance level by changing the length of the prediction window. This suggests that in the early stages, since the available data is limited, choosing smaller windows would help with the performance. However, based on the results in the article we came to know that as more data becomes available, the performance on the longer windows can be significantly improved.

\begin{table}[h]
    \centering
    \caption{This table shows the results of evaluating the trained DWLSTM model in comparison to the best ARIMA models in performing the prediction task on the normalized cumulative death counts due to COVID-19.}
    \label{tab:cumsumpred}
    \begin{tabular}{lcccc}
                                  & \multicolumn{2}{l}{\textbf{DWLSTM (Daily)}}                             & \multicolumn{2}{l}{\textbf{ARIMA* (daily)}}                             \\
\multirow{-2}{*}{}                & \multicolumn{1}{l}{\textbf{Macro}} & \multicolumn{1}{l}{\textbf{Micro}} & \multicolumn{1}{l}{\textbf{Macro}} & \multicolumn{1}{l}{\textbf{Micro}} \\
\textbf{$10$-day window} & $14.68$                           & $33.57$                            & $18.34$                            & $43.38$                            \\
\textbf{$15$-day window}                      & $17.24$                           & $39.98$                            & $22.06$                           & $54.72$                            \\
\textbf{$20$-day window}                      & $24.15$                            & $55.98$                            & $26.62$                            & $67.15 $                           
\end{tabular}
    
\end{table}

\section{Impact of Highly Affected Areas}
As an experiment to show the impact of the highly affected areas in teaching the machine learning model in our approach, we have tried removing the counties of New York state from the dataset and showed the results in Table \ref{nony}. The results indicate that in terms of quantitative assessment, the lack of presence for the highly affected areas causes a significant drop in the loss values. However, the qualitative analysis showed that the models do not perform well in the case of rising values, as the amount of information available on such cases to train the network on is fairly limited. This causes both family of models to be biased in making predictions that tend to underestimate the target values.

\begin{table}[h]
    \centering
    \caption{The performance of DWLSTM and the ARIMA* predictions on the early COVID-19 epidemic (until May 5th, 2020). The objective in training the models was the prediction of normalized cumulative death counts due to the pandemic, and the performance is measured in terms of Daily RMSE on predicting the new daily death counts per county.}
    \label{tab:nony}
    \begin{tabular}{llclc}
                                  & \multicolumn{2}{l}{\textbf{DWLSTM (Daily)}}                      & \multicolumn{2}{l}{\textbf{ARIMA* (daily)}}                      \\
\multirow{-2}{*}{}                & \textbf{Macro}              & \multicolumn{1}{l}{\textbf{Micro}} & \textbf{Macro}              & \multicolumn{1}{l}{\textbf{Micro}} \\ \hline
\textbf{$10$-day window without NY counties}                & \multicolumn{1}{c}{$5.35$}  & $5.91$                             & \multicolumn{1}{c}{$5.07$}  & $5.49$                            \\
\textbf{$10$-day window} & \multicolumn{1}{c}{$14.68$} & $33.57$                          & \multicolumn{1}{c}{$18.33$} & $43.48$                         
\end{tabular}
    
\end{table}


\ifCLASSOPTIONcaptionsoff
  \newpage
\fi



%
\bibliographystyle{IEEEtran}
\bibliography{IEEEabrv,Bibliography}

\begin{thebibliography}{10}
\providecommand{\url}[1]{#1}
\csname url@samestyle\endcsname
\providecommand{\newblock}{\relax}
\providecommand{\bibinfo}[2]{#2}
\providecommand{\BIBentrySTDinterwordspacing}{\spaceskip=0pt\relax}
\providecommand{\BIBentryALTinterwordstretchfactor}{4}
\providecommand{\BIBentryALTinterwordspacing}{\spaceskip=\fontdimen2\font plus
\BIBentryALTinterwordstretchfactor\fontdimen3\font minus
  \fontdimen4\font\relax}
\providecommand{\BIBforeignlanguage}[2]{{%
\expandafter\ifx\csname l@#1\endcsname\relax
\typeout{** WARNING: IEEEtran.bst: No hyphenation pattern has been}%
\typeout{** loaded for the language `#1'. Using the pattern for}%
\typeout{** the default language instead.}%
\else
\language=\csname l@#1\endcsname
\fi
#2}}
\providecommand{\BIBdecl}{\relax}
\BIBdecl

\bibitem{king2012virus}
A.~M. King, M.~J. Adams, E.~B. Carstens, and E.~J. Lefkowitz, ``Virus
  taxonomy,'' \emph{Ninth report of the International Committee on Taxonomy of
  Viruses}, pp. 486--487, 2012.

\bibitem{fan2019bat}
Y.~Fan, K.~Zhao, Z.-L. Shi, and P.~Zhou, ``Bat coronaviruses in china,''
  \emph{Viruses}, vol.~11, no.~3, p. 210, 2019.

\bibitem{liu2019viral}
P.~Liu, W.~Chen, and J.-P. Chen, ``Viral metagenomics revealed sendai virus and
  coronavirus infection of malayan pangolins (manis javanica),''
  \emph{Viruses}, vol.~11, no.~11, p. 979, 2019.

\bibitem{anxietycovid}
C.~of~Disease~Control and Prevention, ``Mental health and coping with stress
  during covid-19 pandemic,''
  \url{https://web.archive.org/web/20200804105944/https://www.cdc.gov/coronavirus/2019-ncov/daily-life-coping/managing-stress-anxiety.html},
  accessed: 2020-08-04.

\bibitem{oliviaweb}
``Olivia health analytics platform,'' \url{http://olivia.cs.ucla.edu},
  accessed: 2020-08-04.

\bibitem{safepaths}
``Private kits: Safepaths; privacy-by-design covid19 solutions using
  gps+bluetooth for citizens and public health officials,''
  \url{https://safepaths.mit.edu }, archived at \url{https://archive.is/zUgQA},
  accessed: 2020-06-05.

\bibitem{raskar2020apps}
R.~Raskar, I.~Schunemann, R.~Barbar, K.~Vilcans, J.~Gray, P.~Vepakomma,
  S.~Kapa, A.~Nuzzo, R.~Gupta, A.~Berke \emph{et~al.}, ``Apps gone rogue:
  Maintaining personal privacy in an epidemic,'' \emph{arXiv preprint
  arXiv:2003.08567}, 2020.

\bibitem{covidapi}
``Covid-19/coronavirus real time updates with credible sources in us and
  canada,'' \url{https://coronavirus.1point3acres.com/en}, archived at
  \url{https://archive.is/J3Vmg}, accessed: 2020-06-05.

\bibitem{yang2020covidnet}
T.~Yang, K.~Shen, S.~He, E.~Li, P.~Sun, L.~Zuo, J.~Hu, Y.~Mo, W.~Zhang, P.~Chen
  \emph{et~al.}, ``Covidnet: To bring the data transparency in era of
  covid-19,'' \emph{arXiv preprint arXiv:2005.10948}, 2020.

\bibitem{kaggle10}
``Novel coronavirus 2019 dataset,''
  \url{https://www.kaggle.com/sudalairajkumar/novel-corona-virus-2019-dataset},
  archived at \url{https://archive.is/zfDjx}, accessed: 2020-06-05.

\bibitem{kaggle9}
``Covid-19 open research dataset challenge (cord-19).''
  \url{https://www.kaggle.com/allen-institute-for-ai/CORD-19-research-challenge},
  archived at \url{https://archive.is/nokIg}, accessed: 2020-06-05.

\bibitem{killeen2020county}
B.~D. Killeen, J.~Y. Wu, K.~Shah, A.~Zapaishchykova, P.~Nikutta, A.~Tamhane,
  S.~Chakraborty, J.~Wei, T.~Gao, M.~Thies \emph{et~al.}, ``A county-level
  dataset for informing the united states' response to covid-19,'' \emph{arXiv
  preprint arXiv:2004.00756}, 2020.

\bibitem{wang2020comparing}
G.~Wang, Z.~Gu, X.~Li, S.~Yu, M.~Kim, Y.~Wang, L.~Gao, and L.~Wang, ``Comparing
  and integrating us covid-19 daily data from multiple sources: A county-level
  dataset with local characteristics,'' \emph{arXiv preprint arXiv:2006.01333},
  2020.

\bibitem{shen2020using}
C.~Shen, A.~Chen, C.~Luo, J.~Zhang, B.~Feng, and W.~Liao, ``Using reports of
  symptoms and diagnoses on social media to predict covid-19 case counts in
  mainland china: Observational infoveillance study,'' \emph{Journal of Medical
  Internet Research}, vol.~22, no.~5, p. e19421, 2020.

\bibitem{pourhomayoun2020predicting}
M.~Pourhomayoun and M.~Shakibi, ``Predicting mortality risk in patients with
  covid-19 using artificial intelligence to help medical decision-making,''
  \emph{medRxiv}, 2020.

\bibitem{wang2020spatiotemporal}
L.~Wang, G.~Wang, L.~Gao, X.~Li, S.~Yu, M.~Kim, Y.~Wang, and Z.~Gu,
  ``Spatiotemporal dynamics, nowcasting and forecasting of covid-19 in the
  united states,'' \emph{arXiv preprint arXiv:2004.14103}, 2020.

\bibitem{covid19simulator}
``Covid-19 simulator,''
  \url{http://web.archive.org/web/20200730193215/https://www.covid19sim.org/},
  accessed: 2020-07-30.

\bibitem{srivastava2020learning}
A.~Srivastava and V.~K. Prasanna, ``Learning to forecast and forecasting to
  learn from the covid-19 pandemic,'' \emph{arXiv preprint arXiv:2004.11372},
  2020.

\bibitem{srivastava2020fast}
A.~Srivastava, T.~Xu, and V.~K. Prasanna, ``Fast and accurate forecasting of
  covid-19 deaths using the sikja model,'' \emph{arXiv preprint
  arXiv:2007.05180}, 2020.

\bibitem{kufel2020arima}
T.~Kufel, ``Arima-based forecasting of the dynamics of confirmed covid-19 cases
  for selected european countries,'' \emph{Equilibrium. Quarterly Journal of
  Economics and Economic Policy}, vol.~15, no.~2, pp. 181--204, 2020.

\bibitem{pei2020initial}
S.~Pei and J.~Shaman, ``Initial simulation of sars-cov2 spread and intervention
  effects in the continental us,'' \emph{medRxiv}, 2020.

\bibitem{kaggle8}
``Us census demographical data,''
  \url{https://www.kaggle.com/muonneutrino/us-census-demographic-data},
  archived at \url{https://archive.is/ZY12v}, accessed: 2020-06-05.

\bibitem{kaggle6}
``Us mortality rates by county,''
  \url{http://ghdx.healthdata.org/record/ihme-data/united-states-mortality-rates-county-1980-2014},
  archived at \url{https://archive.is/juSbk}, accessed: 2020-06-05.

\bibitem{kaggle7}
``Us county-level mortality,''
  \url{https://www.kaggle.com/IHME/us-countylevel-mortality}, archived at
  \url{https://archive.is/xEVs3}, accessed: 2020-06-05.

\bibitem{dwyer2016us}
L.~Dwyer-Lindgren, A.~Bertozzi-Villa, R.~W. Stubbs, C.~Morozoff, M.~J. Kutz,
  C.~Huynh, R.~M. Barber, K.~A. Shackelford, J.~P. Mackenbach, F.~J. van Lenthe
  \emph{et~al.}, ``Us county-level trends in mortality rates for major causes
  of death, 1980-2014,'' \emph{Jama}, vol. 316, no.~22, pp. 2385--2401, 2016.

\bibitem{kaggle4}
``Diversity index of us counties,''
  \url{https://www.kaggle.com/mikejohnsonjr/us-counties-diversity-index},
  archived at \url{https://archive.is/uX9iX}, accessed: 2020-06-05.

\bibitem{kaggle432}
``Us drought monitor,'' \url{https://droughtmonitor.unl.edu/}, archived at
  \url{https://archive.is/P76Bb}, accessed: 2020-06-05.

\bibitem{kaggle3}
``United states droughts by county,''
  \url{https://www.kaggle.com/us-drought-monitor/united-states-droughts-by-county},
  archived at \url{https://archive.is/JgGoj}, accessed: 2020-06-05.

\bibitem{electiondata}
``County presidential election returns2000-2016, 2018,''
  \url{https://doi.org/10.7910/DVN/VOQCHQ}, archived at
  \url{https://archive.is/cLVL5}, accessed: 2020-06-05.

\bibitem{kaggle2}
``Icu beds by county in the us,''
  \url{https://www.kaggle.com/jaimeblasco/icu-beds-by-county-in-the-uss},
  archived at \url{https://archive.is/QgAoO}, accessed: 2020-06-05.

\bibitem{kaggle1}
``Us household income statistics,''
  \url{https://www.kaggle.com/goldenoakresearch/us-household-income-stats-geo-locations},
  archived at \url{https://archive.is/iJaLT}, accessed: 2020-06-05.

\bibitem{cdccovidushospitalization}
CDC, ``A weekly summary of us covid-19 hospitalization data,''
  \url{https://gis.cdc.gov/grasp/COVIDNet/COVID19\_1.html}, archived at
  \url{https://archive.is/qs0IJ}, accessed: 2020-06-05.

\bibitem{cdccovidushospitalizationlabconfirmed}
------, ``Laboraty-confirmed covid-19 associated hospitalizations,''
  \url{https://gis.cdc.gov/grasp/covidnet/COVID19\_3.html}, archived at
  \url{https://archive.is/Mw9d1}, accessed: 2020-06-05.

\bibitem{nsrestaurant}
N.~R. Association, ``State statistics,''
  \url{http://web.archive.org/web/*/https://www.restaurant.org/research/state},
  accessed: 2020-07-15.

\bibitem{healthdata}
``Us data for download,''
  \url{http://web.archive.org/web/20200121125528/http://www.healthdata.org/us-health/data-download},
  accessed: 2020-04-02.

\bibitem{wold1987principal}
S.~Wold, K.~Esbensen, and P.~Geladi, ``Principal component analysis,''
  \emph{Chemometrics and intelligent laboratory systems}, vol.~2, no. 1-3, pp.
  37--52, 1987.

\bibitem{tishby2015deep}
N.~Tishby and N.~Zaslavsky, ``Deep learning and the information bottleneck
  principle,'' in \emph{2015 IEEE Information Theory Workshop (ITW)}.\hskip 1em
  plus 0.5em minus 0.4em\relax IEEE, 2015, pp. 1--5.

\bibitem{fourreasonsblacksaffected}
E.~Scott, ``4 reasons coronavirus is hitting black communities so hard,''
  \url{http://web.archive.org/web/20200728222505/
  https://www.washingtonpost.com/politics/2020/04/10/4-reasons-coronavirus-is-hitting-black-communities-so-hard/},
  accessed: 2020-07-28.

\bibitem{nytimesdataset}
``Covid-19 data in the united states,''
  \url{https://usafacts.org/visualizations/coronavirus-covid-19-spread-map},
  archived at \url{https://archive.is/WefdJ}, accessed: 2020-05-10.

\bibitem{usfacts}
``Us facts dataset,''
  \url{https://usafacts.org/visualizations/coronavirus-covid-19-spread-map},
  archived at \url{https://archive.is/tt0ih}, accessed: 2020-07-30.

\end{thebibliography}

%








\end{document}